# Squirrel: Testing Database Management Systems with Language Validity and Coverage Feedback

Rui Zhong[†][∗], Yongheng Chen[§][∗], Hong Hu[§], Hangfan Zhang[†], Wenke Lee[§] and Dinghao Wu[†]

[†]*Penn State University*  [§]*Georgia Institute of Technology*

## Abstract

Fuzzing is an increasingly popular technique for verifying software functionalities and finding security vulnerabilities. However, current mutation-based fuzzers cannot effectively test database management systems (DBMSs), which strictly check inputs for valid syntax and semantics. Generation-based testing can guarantee the syntax correctness of the inputs, but it does not utilize any feedback, like code coverage, to guide the path exploration.

In this paper, we develop Squirrel, a novel fuzzing framework that considers both language validity and coverage feedback to test DBMSs. We design an intermediate representation (IR) to maintain SQL queries in a structural and informative manner. To generate syntactically correct queries, we perform type-based mutations on IR, including statement insertion, deletion and replacement. To mitigate semantic errors, we analyze each IR to identify the logical dependencies between arguments, and generate queries that satisfy these dependencies. We evaluated Squirrel on four popular DBMSs: SQLite, MySQL, PostgreSQL and MariaDB. Squirrel found 51 bugs in SQLite, 7 in MySQL and 5 in MariaDB. 52 of the bugs are fixed with 12 CVEs assigned. In our experiment, Squirrel achieves 2.4×-243.9× higher semantic correctness than state-of-the-art fuzzers, and explores 2.0×-10.9× more new edges than mutation-based tools. These results show that Squirrel is effective in finding memory errors in database management systems.

## 1 Introduction

Database Management Systems (DBMSs) are the integral components of modern data-intensive systems [9, 10, 13, 29, 39, 58]. Like all other complicated systems, DBMSs contain many bugs that not only affect their functionalities but also enable malicious attacks. Among all the threats, the infamous memory error bugs enable attackers to leak and even corrupt memory of running DBMS processes, which may finally lead to remote code execution [28, 38], database breach [7, 30] or denial-of-service (DoS) [16, 24]. For example, the recent "Collection #1" data breach reveals 773 million email addresses and 21 billion passwords [36].

Generation-based testing techniques are the *de facto* bug-finding tools for DBMSs [5]. These techniques require developers to create a formal model that precisely captures the definition of SQL (Structured Query Language). Based on the model, the tools enumerate all possible SQL queries to verify the functionalities of DBMSs or find bugs. However, generation-based testing tools have limited effectiveness as they distribute the effort evenly to every SQL query.

Considering the infinite input space and rare bug-triggering queries, this bruteforce-like enumeration is ineffective in finding memory error bugs from DBMSs.

In recent years, fuzzing has been widely adopted as a software testing technique to detect memory error vulnerabilities [33, 43, 46, 66]. Different from generation-based testing, fuzzing relies on the random mutation to create new test cases, and utilizes feedback, like code coverage, to guide the input space exploration. Starting from the seed corpus, a fuzzer randomly mutates existing inputs, like flipping several bits, to generate slightly different variants. It runs the target program with the new input and detects abnormal behaviors such as execution crashes and assertion failures. During the execution, the fuzzer also records the code path information, like executed blocks or branches. The input that triggers new code paths has a higher priority to be selected for another round of mutation. With a large amount of effort spent on improving the fuzzing efficiency [34, 49, 63] and efficacy [26, 31, 42, 51], fuzzers have successfully found thousands of bugs from popular applications [55].

However, it is challenging to apply fuzzing techniques to test DBMSs, as DBMSs perform two correctness checks, the *syntactic check* and the *semantic check* before executing an SQL query. Specifically, the DBMS first parses each SQL query to get its syntactic meaning. If the query has any grammar errors, the DBMS will stop the execution and immediately bail out with an error message. Otherwise, the DBMS further checks the query for semantic errors, like using a non-existent table, and will bail out in any case of semantic errors. After these two checks, the DBMS creates several execution plans and picks the most efficient one to execute the query. Therefore, to reach the deep logic of a DBMS, the query should be correct syntactically and semantically.

Random byte mutation used by current fuzzing techniques cannot easily generate syntax-correct or semantics-correct inputs. For example, AFL, the representative mutation-based fuzzer [66], can generate 20 million queries for SQLite [4] within 24 hours, but only 30% of them can pass the syntax check, and 4% have correct semantic meaning. However, most of the DBMS code is responsible for query plan construction, optimization, and execution, and only a small portion is used for syntax-check and semantics-check. For example, 20,000 semantics-correct queries generated by AFL trigger 19,291 code branches in SQLite, while the same number of syntax-incorrect queries only cover 9,809 branches – only 50.8% of the former. Therefore, current fuzzing techniques cannot trigger the deep logic of DBMSs nor comprehensively explore program states.

In this paper, we propose a system, Squirrel, to address these challenges so that we can effectively fuzz DBMSs. Our system embeds two key techniques, the *syntax-preserving mutation* and the *semantics-guided instantiation*. To generate syntax-correct SQL queries, we design an intermediate representation (IR) to maintain queries in a structural and informative manner. Each IR statement contains at most two operands, and each operand is also another IR statement. Each IR has a structure type that indicates the syntactic structure (e.g., SELECT a FROM b), and data types (e.g., table name). Before the mutation, our system strips concrete data from the IR and only keeps a skeleton of operations. Then, we perform three random mutations, including inserting type-matched operands, deleting optional operands or replacing operands with other type-matched ones. The type-based mutation guarantees the generated query has the correct syntax. Next, we perform the query analysis to figure out the expected dependencies between different IR data. For example, the data in a SELECT clause should be a column of the table in the FROM clause. We fill each stripped IR with concrete data, and make sure they satisfy all expected dependencies. At last, we translate the IR back into SQL and feed it to the DBMS for testing. SQUIRREL combines the benefits of the coverage-based fuzzing (i.e., guided exploration) and the model-based generation (i.e., high language correctness), and thus can trigger the deep logic of DBMSs and find severe bugs.

We implement SQUIRREL with 43,783 lines of C++ code, which mainly focuses on the syntax-preserving mutation and the semantics-guided instantiation. We reuse AFL's code for coverage collection and input prioritization. Our design of SQUIRREL is generic and it should work with other fuzzers after some engineering effort.

To understand the effectiveness of our system, we use SQUIRREL to test four popular DBMSs: SQLite, MySQL, PostgreSQL and MariaDB. SQUIRREL successfully found 63 memory error issues within 40 days, including 51 bugs in SQLite, 7 bugs in MySQL and 5 bugs in MariaDB. As a comparison, Google OSS-Fuzz detected 19 bugs from SQLite in 40 months and 15 bugs from MySQL in 5 months [34]. We have responsibly reported all of these bugs to the DBMS developers and received positive feedback. At the time of paper writing, 52 bugs have been fixed. We even get CVE numbers for 12 bugs due to their severe security consequences, like stealing database contents.

We inspect various aspects of fuzzing, and compare SQUIRREL with other state-of-the-art tools, including the mutation-based fuzzer AFL and Angora, the generation-based tool SQLsmith, the structural fuzzer GRIMOIRE and the hybrid fuzzer QSYM. During the 24-hour testing, SQUIRREL successfully finds nine unique bugs while others detect one or zero bugs. SQUIRREL discovers 2.0×-10.9× more new edges than mutation-based tools, and achieves a comparable result to the generation-based tester SQLsmith. It also gets 2.4×-243.9× higher semantic correctness than other tools.

We make the following contributions in this paper:

- We propose syntax-preserving mutation and semantics-guided instantiation to address the challenges of fuzzing DBMSs.
- We implement SQUIRREL, an end-to-end system that combines mutation and generation to detect DBMS bugs.
- We evaluated SQUIRREL on real-world DBMSs and identified 63 memory error issues. The result shows that SQUIRREL outperforms existing tools on finding bugs from DBMSs.

We plan to release the code of SQUIRREL to help DBMS developers test their products and to boost future research in securing DBMSs.

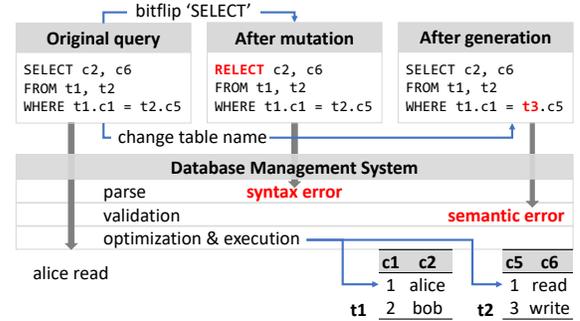

**Figure 1: Challenges of testing DBMSs.** A DBMS takes four steps to process one SQL query. Among them, *parse* checks syntactic correctness, and *validation* examines semantic validity. Random mutation unlikely guarantees the syntactic correctness, while grammar-based generation may fail to enforce semantic correctness.

## 2 Problem Definition

In this section, we first briefly describe how a DBMS handles SQL queries. Then, we introduce existing DBMS testing techniques and illustrate their limitations in finding bugs hidden in the deep logic. Finally, we present our insight to solve this problem.

### 2.1 Query Processing in DBMS

Modern DBMSs usually process an SQL query in four phases: parse, validation, optimization and execution [6], as shown in Figure 1. After receiving an SQL query, the DBMS first parses the query to get its syntactic meaning. The parser breaks the query into individual tokens, and checks them against the grammar rules. If any syntactic error is detected, the DBMS will immediately terminate the execution and return an error message to the client. Second, the DBMS checks the semantic correctness of the query in the validation phase, like whether tables exist in the database or columns are unambiguous. Most semantic errors can be detected in this phase. Next, in the optimization phase, the query optimizer constructs several possible query plans and attempts to determine the most efficient one for executing the given query. Finally, the DBMS executes the chosen plan on the database and sends the result back to the client. Therefore, the execution will reach the second phase only if the query is *syntactically* correct, and will proceed to the last two phrases if the query is *semantically* correct.

**Motivating Example.** The "Original query" in Figure 1 first joins two tables t1 and t2, and searches for the rows where the c1 column of t1 is the same as the c5 column of t2. For each matched row, the query returns the value of c2 and c6. The DBMS finds that this query passes the syntactic check and the semantic check. It searches in the database and finally returns "alice read".

### 2.2 Challenges of DBMS Testing

There are mainly two ways to generate SQL queries for testing DBMSs: model-based generation and random mutation. The model-based generation follows a precise grammar-model and thus can construct syntactically correct inputs. For example, SQLsmith [5], a popular DBMS testing tool, generates syntax-correct test cases from the abstract syntax tree (AST) [60] directly. However, without any guidance, the model-based generation sequentially scans the whole input space. Considering that many queries are handled in the same way by DBMSs, this method cannot efficiently explore the program's state space. Further, the generation-based method can hardly guarantee the semantic correctness [35], and queries with incorrect semantics will be filtered out by the DBMS during the validation. Figure 1 shows a query constructed by a generator (After generation). Although this query is syntactically correct, it cannot be executed because the table t3 in the WHERE clause does not exist in the current database.

Random mutation updates existing inputs to generate new ones. To improve the performance, fuzzers utilize the feedback from previous executions to evaluate the priority of the generated inputs. If the feedback indicates the previous input is interesting, like triggering a new execution path, fuzzers will put it in a queue for further mutation. In this way, fuzzers will collect more and more interesting test cases and thus can explore the program's state space efficiently. Statistics show that random mutation with feedback-driven works well in many software. For example, Google found over 5,000 vulnerabilities with their feedback-driven mutation-based fuzzer [43, 55]. However, grammar-blind mutation strategies have low effectiveness in handling structured inputs, like SQL and JavaScript [60]. For example, random flipping the bits of a SQL keyword hardly produces another valid keyword, and the whole query will become syntactically incorrect. Figure 1 shows such a case (After mutation), where flipping the last bit of S in SELECT leads to an invalid keyword RELECT. The DBMS will reject the new query in the parse phase.

We design evaluation to understand the quality of AFL-generated SQL queries, and the importance of syntax-correctness and semantics-correctness. Specifically, we use AFL to test SQLite for 24 hours, which generates 20 million queries. However, only about 30% of them are syntactically correct, and merely 4% for them can pass semantic checks. We randomly pick 20,000 semantics-correct queries, and find that they trigger 19,291 distinct code branches in SQLite. The same number of syntax-incorrect and semantics-incorrect queries only reach 9,809 and 12,811 branches, respectively. These results show the low validation rate of AFL-generated queries, and the importance of semantics-correctness for exploring the program state space.

## 2.3 Our Approach

Our idea in this paper is to introduce syntax-correct and semantics-aware mutation into fuzzing, so we can take advantage of mutation-based techniques and generation-based mechanisms to maximize our ability to test DBMSs.

**Generating Syntax-Correct Queries.** We design a new intermediate representation (IR) to maintain SQL queries in a structural and informative way and adopt type-based mutations to guarantee the syntactic correctness. Each IR statement simply contains at most two operands, and therefore our mutation just has to handle two values. Each statement has an associated grammar type, like SelectStmt for SELECT statement, while each data has a semantic type, like table name. Our mutation performs type-based operations, including inserting type-matched operands, deleting optional operands and replacing operands with type-matched ones. We strip the concrete data from each IR, like table names, to focus on mutating the skeleton. The IR-based mutation effectively preserves the syntactic correctness. Some generation-based tools generate SQL queries from the AST. However, due to the strict type-constraint and complicated operations, mutating AST is as challenging as modifying SQL queries.

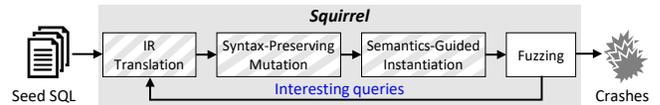

**Figure 2: Overview of Squirrel.** Squirrel aims to find queries that crash the DBMS. Squirrel first lifts queries from SQL to IR; then, it mutates IR to generate new skeletons; next, it fills the skeleton with concrete operands; finally, it runs the new query and detects bugs.

**Improving Semantic Correctness.** Since ensuring the semantic correctness of generated SQL queries is proved to be NP-hard [44], we will try practical solutions to improve the semantic correctness as much as possible. Existing generation-based tools define a set of query templates. Each template represents a complete query and contains specific, static constraints between operands [20]. However, due to the limited human effort, these frameworks cannot guarantee the expressiveness of their SQL templates. We tackle this problem through dynamic *query instantiation*. Given the skeleton of a syntax-correct SQL query (i.e., without concrete operands), our method first builds its data dependency graph according to predefined basic rules. For example, the operand of SELECT can be a column name of the table used in FROM. Then, we try to fill the skeleton with concrete operands whose relations satisfy the data dependency graph. With the instantiation, the semantic correctness rate is high enough for testing DBMSs.

## 3 Overview of Squirrel

Figure 2 shows an overview of our DBMS testing framework, Squirrel. Given a set of normal SQL queries, Squirrel aims to find queries that render the execution of DBMSs crash. A query means a test case and may contain multiple SQL statements. Squirrel starts with an empty database and requires the query to create the content. Squirrel achieves its goal with four key components: Translator, Mutator, Instantiator, and SQL Fuzzer. First, Squirrel selects one query $I$ from a queue that consists of both initial queries and saved interesting queries. Second, the Translator translates $I$ into a vector of IRs $V$. Meanwhile, the Translator strips the concrete values from $V$ to make it a query skeleton. Our Mutator modifies $V$ through insertion, deletion and replacement to produce a new IR vector $V'$ — $V'$ is syntactically correct. Next, our Instantiator performs data dependency analysis of $V'$ and builds a data dependency graph. Then, the Instantiator selects new concrete values that satisfy the data dependency and fills $V'$ with these values. Since the data dependency is satisfied, $V'$ is likely to be semantically correct. Finally, we convert $V'$ back to a SQL query $I'$ and run the DBMS with $I'$. If the execution crashes, we find an input that triggers a bug. Otherwise, if $I'$ triggers a new execution path of the program, we save it into the queue for further mutation.

## 4 Intermediate Representation

We design an intermediate representation (IR) of SQL to support the syntax-correct query mutation. We translate each query from SQL to our IR, mutate the IR and translate the new IR back to SQL

```
1  // l: left child, r: right child, d: data, t: data type
2  V1 = (Column,      l=0,  r=0,   op=0, d="c2", t=ColumnName);
3  V2 = (ColumnRef,   l=V1, r=0,   op=0, d=0);
4  V3 = (Expr,        l=V2, r=0,   op=0, d=0);
5  V4 = (Column,      l=0,  r=0,   op=0, d="c6", t=ColumnName);
6  V5 = (ColumnRef,   l=V4, r=0,   op=0, d=0);
7  V6 = (Expr,        l=V5, r=0,   op=0, d=0);
8  V7 = (SelectList,  l=V3, r=V6,  op=0, d=0);
9  // the optional left child can be DINSTRICT
10 V8 = (SelectClause, l=0, r=V6, op.prefix="SELECT", d=0);
11 ...
12 //Unknown type for intermediate IRs
13 Va = (Unknown,     l=V8, r=V14, op=0, d=0);
14 Vb = (Unknown,     l=Va, r=V25, op=0, d=0);
15 // the optional right child can be an ORDER clause
16 V26 = (SelectStmt, l=Vb, r=0,   op=0, d=0);
```

**Figure 3: IR of the running example SQL query.** The corresponding AST tree is shown in Figure 4.

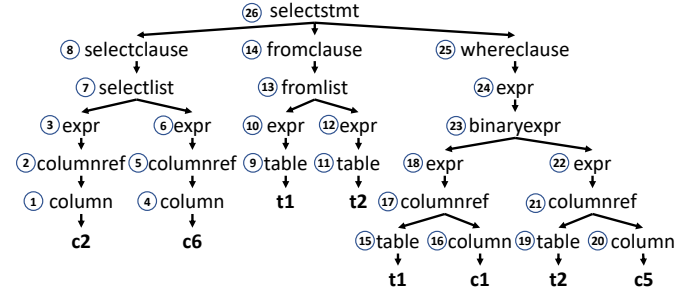

**Figure 4: AST of the running example.** SQUIRREL parses the SQL query and represents it in AST, and finally translate AST to IR.

query for execution. Our design of the IR aims to achieve three goals: the IRs can represent any SQL statements (**expressive**); the format and operation of IRs are uniform (**general**); the translation between IR and SQL is efficient (**simple**).

The IR is in the static single assignment (SSA) form. A query, or a test case, contains one or more IR statements. Each statement is an assignment, where the left-hand side is the destination variable and the right-hand side is either a literal or an operator with operands. We add the following fields in IR to store the necessary information.

- `ir_type`: the type of one IR statement. This type is based on the corresponding node in the AST, like column type for column names or expr type for expressions. We also define a special type Unknown to represent intermediate statements that have no corresponding node in the AST.
- `operator`: consisting of SQL keywords [11] and mathematical operators [12]. It indicates the operation the IR performs and includes three parts: the prefix `op_prefix`, the interfix `op_mid` and the suffix `op_suffix`. For example, the IR of "CREATE trigger BEGIN list END" has prefix CREATE, interfix BEGIN and suffix END.
- `left_operand`, `right_operand`: the operands of the IR operator. The operand is either another IR statement, or can be NULL if the operand is optional or not required.
- `data_value`: the concrete data the IR carries, like table name t1.
- `data_type`: data type, like ColumnName for column names.

We provide the formal definition of the IR grammar in Appendix A.

Figure 3 shows the IRs of our motivating example in Figure 1 (Original query). The corresponding AST is given in Figure 4. V1 and V4 represent the column names c2 and c6, which are corresponding to nodes ① and ④ in Figure 4. They do not contain any operator or operand but have the ColumnName data type and proper data values. V2 and V5 define references to columns (V1 and V4), and V3 and V6 create two expressions. Each of them only has one operand. V7 describes the parameter list of SELECT, including c2 and c6. V8 represents the SELECT clause, which could have DISTINCT as its left operand (NULL here). SELECT appears before the left operand, so it is the operator prefix. Since our IR only allows at most two operands, we have to use two intermediate nodes, Va and Vb, to connect three nodes ⑧, ⑭ and ㉕ to construct the IR of SelectStmt. Their `ir_types` are set to Unknown. Finally, V26 defines the SELECT statement, which is node ㉖ in Figure 4.

Our IR is just a sequence of assignment statements. This linear representation, different from tree or graph structures (like AST),

helps developers to adopt unified and simple mutation strategies. We can perform statement insertion, deletion and replacement while keeping the syntactic correctness. We present the algorithms about translation between SQL queries and IRs in Appendix B.

## 5 Syntax-Preserving Mutation

We classify tokens in an SQL query into two groups based on their functionalities. SQL keywords and mathematical operators define what operations to be performed, and we call these tokens *structure*. Other tokens specify the targets of defined operations, and we call them *data*. Data can be literal that makes basic sense, like a constant value 1, or can express semantic meaning, like table names.

We observe that changing structure tokens has more impact on the DBMS execution than that of changing data tokens. The difference comes from two reasons. First, altering structure will change the operations of the query, and thus trigger different functions, while a DBMS may use the same logic to handle different literal data. For example, SQLite takes almost the same path to process query A:"SELECT c FROM t WHERE c=1" and query B:"SELECT c FROM t WHERE c=10", but uses significantly different code to handle C:"SELECT c FROM t WHERE c>1". Second, randomly modifying semantic-related data is likely to generate a semantically incorrect query, which a DBMS will refuse to execute. For example, replacing c in query A with the column in another table leads to an invalid query. In either case, random data mutation is less productive than random structure mutation.

Therefore, we strip data from the query IR and apply mutation mainly on structures. We leave the data modification in §6.

### 5.1 Structure-Data Separation

We walk through the IRs to replace each data with a predefined value based on its type `data_type`. Specifically, we replace semantic data with string **"x"**, change constant numbers to 1 or 1.0 and update all strings to **"a"**. Therefore, after the separation, the running example "SELECT c2,c6 FROM t1,t2 WHERE t1.c1=t2.c5" becomes "SELECT x,x FROM x,x, WHERE x.x=x.x". Both query A and B become "SELECT x FROM x WHERE x=1", while query C is changed to "SELECT x FROM x WHERE x>1".

**Storing IR in Library.** We use a dictionary called *IR library* to store various IRs. The key of the dictionary is the IR type, while the value is a list of IRs. IRs in one list have the same type, and their structures are exclusively different. For example, after separation, queries A, B, and C have the same SelectStmt type, and they should be stored in the same list. We remove query B from the list as it shares the same structure as A. Whenever we need an IR of a certain type, we find the corresponding list from the dictionary and randomly return one element. As we show in Figure 2, Squirrel accepts seed queries to initialize the IR library. Whenever Squirrel finds that the generated IR has a new structure, we add it to the corresponding list in the library. We set a limit on the maximum number of IRs in the library to avoid excessive memory usage.

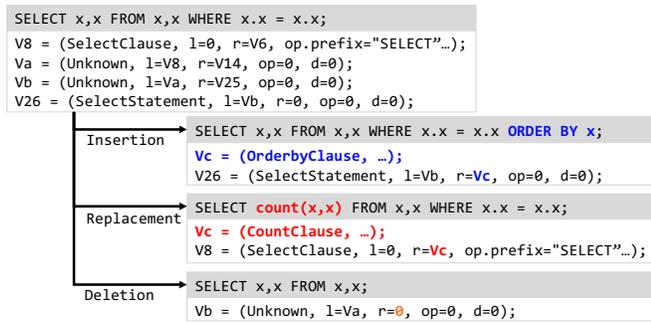

Figure 5: Mutation strategies on IR programs, including type-based insertion and replacement, and deletion of optional operands.

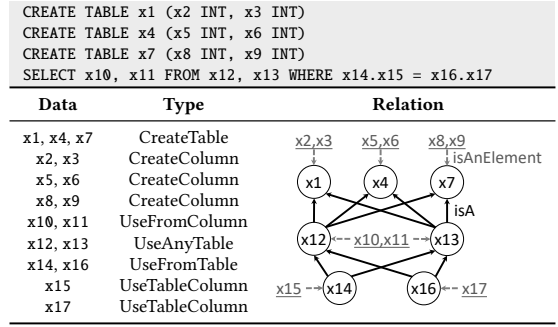

Figure 6: Data dependency example. This example consists of three new CREATE statements and our running example. In "Relation", we show two types of relations: "isAnElement" (dashed line) and "isA" (solid line).

## 5.2 Type-Based Mutation

We define a set of type-based mutations to update the left and right operands of an IR or the IR itself. Our mutation focus on operands as other members of the IR cannot be easily changed: the operator is closely related to the IR type, like the SELECT operator in SelectClause IR, while data_type is decided by its position in the query, like a variable after "CREATE TABLE" must be a table name. Therefore, our mutations either operate on the whole IR or modify its operands. Specifically, for each IR $v$ in the IR program we perform the following mutations with certain probability:

- **Insertion** adds an IR into an appropriate position of $v$. If the left child of $v$ is empty, we randomly pick an IR $w$ from the IR library that shares the same type as $v$. If the left child of $w$ is not empty, we use it as the left child of $v$. The same operation applies to the right child.
- **Replacement** changes $v$ or its operands. We first randomly pick one IR $w$ of the same type as $v$ from the IR library. Then we copy the children of $w$ to $v$, or we can replace $v$ with $w$ and update all $v$'s references to $w$.
- **Deletion** removes a $v$ as a whole by simply replacing it with an empty IR. The same operation can be applied to its children.

Since we essentially manipulate IRs based on their types, the syntactic correctness is preserved with a high probability. To further improve the syntactic correctness, we convert the mutated IRs back to an SQL query and perform syntax validation with our parser. If the parsing succeeds, we conclude that this query has no syntax errors and will use it in the next stage. Otherwise, we discard the new IRs and try to generate another one.

Figure 5 shows an example of mutating the IRs of our running example to generate three new queries. Specifically, we insert an ORDERBY clause to the right child of V26; we replace the right child of V8 with a CountClause, where the new query counts the rows of the original results; we delete the right child of Vb to effectively remove the WHERE clause. All three new IRs are syntactically correct.

**Unknown Type.** As we mention in §4, some IRs have type Unknown as they do not have corresponding nodes in the AST. We use Unknown type to perform fuzzy type-matching without searching for concrete types, which may accelerate our query generation. The syntax validation, which always needs one-time parsing, is unaffected. However, without accurate type-matching, Squirrel may create some invalid queries.

## 6 Semantics-Guided Instantiation

Semantics-correct queries enable fuzzers to dig deeply into DBMSs' execution logic and discover bugs effectively. However, generating semantics-correct test cases is an unsolved challenge for fuzzing programs that take structured, semantics-binding inputs [44]. Previous research shows that 90% of the test cases generated by js-funfuzz [48], a state-of-the-art JavaScript fuzzer, are semantically invalid [35]. Similar problems also exist in DBMS testing.

We propose a data instantiation algorithm to improve the semantic correctness of generated SQL queries. As mentioned in §5, after mutation the IR program is a syntax-correct skeleton with data stripped. Our instantiator first analyzes the dependency between different data, and fill the skeleton with concrete values that satisfy all dependencies. After the instantiation, the query has a high chance to be semantically correct.

### 6.1 Data Dependency Inference

Data dependency describes the relationship between semantic-binding data. Any unsatisfied dependency will make the query fail semantic checks. Figure 6 shows the data dependencies among four SQL statements, including three CREATE statements and our motivating example. Our syntax-correct mutation has replaced each variable with x. To distinguish different x, we assign an index to each of them. These four statements contain two types of relationships: one defines A is an element of B (isAnElement), shown as gray-dashed lines like x2 is a column of x1; another describes that A can be B (isA), shown as black-solid lines like x12 can be x1.

We define a set of rules to automatically infer dependencies between data in the query. These rules follow two principles. The *lifetime* principle requires us to create SQL variables before using them, and stop using a variable after its deletion. The *customization*

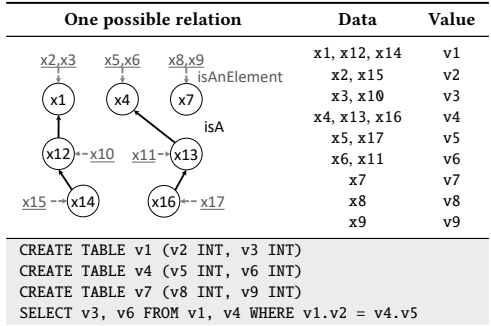

| One possible relation | Data | Value |
|---|---|---|
| (graph structure) | x1, x12, x14 | v1 |
|  | x2, x15 | v2 |
|  | x3, x10 | v3 |
|  | x4, x13, x16 | v4 |
|  | x5, x17 | v5 |
|  | x6, x11 | v6 |
|  | x7 | v7 |
|  | x8 | v8 |
|  | x9 | v9 |

```
CREATE TABLE v1 (v2 INT, v3 INT)
CREATE TABLE v4 (v5 INT, v6 INT)
CREATE TABLE v7 (v8 INT, v9 INT)
SELECT v3, v6 FROM v1, v4 WHERE v1.v2 = v4.v5
```

**Figure 7: Instantiation of IR structure.** We create one concrete dependency graph from the dependencies of Figure 6, replace all place-holder x and finally get one concrete new query.

principle requires us to consider the data types, scopes and operations to determine the relationship. We need to refine data types mentioned in §4 to accurately describe data dependencies.

**Data Type.** We refine each data type so that it not only describes the semantic meaning but also reflects the usage context. Database elements in different statements, even with the same basic type, can have different dependencies. Based on the lifetime principle, we include the define/use information into the data type, to indicate whether the element is a new definition or use of an existing one. For example, a table in a `CREATE TABLE` clause will have `CreateTable` type, while the table in a `FROM` clause will have `UseTable` type. Based on the customization principle, we also include the scope of the data to show where to find the potential candidate values. For example, a table in a `FROM` clause can be any defined table, thus having type `UseAnyTable`, while the table in a `WHERE` clause must be one of the tables in the `FROM` clause, thus having the `UseFromTable` type. The refined data type of a variable is determined by its position in the AST. Therefore, SQUIRREL identifies and sets the data types during the query parse and translation.

Figure 6 shows the refined type for each IR data. For example, x1 is a newly defined table and thus has type `CreateTable`. x2 and x3 have type `CreateColumn`. x12 has type `UseAnyTable` as it can be any defined table (x1, x4, x7), while x14 can only be tables listed in the `FROM` clause. x10 can be any column of tables in `FROM`, while x15 can only be the column of table x14.

**Data Relation Rule.** With refined data types, we can further define data relation rules that can help automatically infer data dependencies. A relation rule is a tuple $(\alpha, \beta, \gamma, S)$ of four elements: $\alpha$ is the relation target and $\beta$ is the relation source; $\gamma$ defines the relationship; $S$ denotes the scope of the relationship, including `intraStmt` for relations in the same statement, `interStmt` for relations across multiple statements, `any` for any instance while `nearest` for the element with the shortest path according to the Define-Use chain. We define eight general rules for all DBMSs, one extra rule for `SQLite`, two extra rules for `MySQL`, two extra for `MariaDB` and one for `PostgreSQL`. For example, the relation rule (`UseFromTable`, `UseTableColumn`, `isAnElement`, `nearest`) means the data of type `UseTableColumn` is an element of the nearest data within the same statement that has type `UseFromTable`. In Figure 6, we can infer the relationship between x15 and x14 with this relation rule.

**Algorithm 1:** Semantic Instantiation

**Input:** *graph*: The dependency graph of the IR program
**Output:** *sqlQuery*: A SQL query that can be executed by DBMS. Empty if error

1 **Procedure** Instantiation(*graph*)
2     dataMap ← map();  relationMap ← map();
3     **for** *each tree T ∈ graph* **do**
4         **for** *each node N ∈ T* **do**    // in BFS, statement order
5             **if** *N.data is literal* **then** N.data ← Random predefined or generated data ;
6             **else if** *N has no parent* **then**
7                 **if** *N.datatype is Definition* **then**
8                     N.data ← GenerateUniqueString();
9                     dataMap[N.datatype].insert(N.data) ;
10                 **else if** *N.datatype is Use* **then**
                    // Use before definition
11                     **if** *dataMap[N.datatype] is empty* **then return** Null ;
                    // Some predefine name, like Function name
12                     **else** N.Data ← anyone ∈ dataMap[N.datatype] ;
13                 **else return** Null    // Delete before definition ;
14             **else**
15                 pN ← The parent node of N;
16                 **if** *N.datatype is Definition* **then**
17                     N.data ← GenerateUniqueString();
18                     dataMap[N.datatype].insert(N.data) ;
19                     relationMap[pN.data].insert(N.data);
20                 **else if** *N.datatype is Use* **then**
21                     **if** *N.type is the same as pN.type* **then** N.data ← pN.data; ;
22                     **else if** *relationMap[pN.data] is not empty* **then**
23                         N.data ← anyone ∈ relationMap[pN.data];
24                     **else return** Null ;
25                 **else**
26                   N.Data ← pN.data;
27                 Delete pN.data from dataMap and relationMap;
28     **return** sqlQuery ← Tranlate the instantiated IR program to a SQL query;

**Dependency Graph.** With data types and relations, SQUIRREL automatically constructs a dependency graph $G = \{V, E\}$ for each mutated IR program. Each node in $V$ is an IR data and its data type. Each edge in $E$ describes one relationship from the edge source to the target. If a data type potentially depends on two or more data types, we randomly choose one to avoid circular dependency. Also, if there are multiple candidate values for the dependent type, SQUIRREL randomly picks one to establish the edge. In this way, every node in the graph has at most one parent, and the dependency graph is formed as one tree or several trees. Appendix C includes more details on the dependency graph construction.

Figure 7 shows one possible dependency graph constructed from Figure 6. For example, we choose x1 as the dependency of x12, although it can be any one of (x1, x4, x7) according to Figure 6. Based on different choices, we can create multiple concrete data dependency graphs for each mutated IR. There are some details not shown in the figure, like x10 should be one of x2 and x3. SQUIRREL handles these implicit dependencies properly.

### 6.2 IR Instantiation

SQUIRREL instantiates the query by filling in concrete data. Algorithm 1 shows our instantiation algorithm. For each tree in the dependency graph, we sort its nodes based on the breadth-first search and the statement order, which guarantees the lifetime correctness. For literal data like integers, we set it to a random value or one from a predefined value set (line 5). For semantic-binding data, we fill in appropriate, valid names. During the process, we maintain two maps: dataMap tracks unique names with different types, while relationMap maps each element to its dependency. If the current node has no dependency (line 6), it either defines a new variable, where we create a new unique string as its name (line 7-9), or is a predefined term, like a function name (line 10-13). If the current node has a parent, we know that it has some dependencies (line 15-27): if the current node creates a new variable, we simply generate a unique string for it (line 16-19); if the current node uses a variable, we check the relationMap to find a proper value for it (line 22-23). Finally, we translate the IR program back into an SQL query and return. If the process fails because of unsatisfiable dependency, the IR program will contain semantic errors.

Table 1: Code size of Squirrel components, totally 43,783 LoC.

| Module | Translator | Mutator | Instantiator | Fuzzer | Others |
|---|---|---|---|---|---|
| Language | C++/Bison/Flex | C++ | C++ | C++ | C++/Make |
| LoC | 32,947 | 4,572 | 1,880 | 2,208 | 2,176 |

Figure 7 shows the result of instantiation the in Data and Value column, and the final SQL query. Squirrel assigns v1 to x1 as it is a CreateTable without dependency. Due to the statement order, we process x2 and x3 next and allocate names v2 and v3 for them, respectively. We handle x4-x9 in a similar way. For x12, it is with type UseAnyTable and has a dependency on x1, so based on line 23 of Algorithm 1, we assign x1's name v1 to x12. Other data can be instantiated in the same way.

## 7 Implementation

We implement Squirrel with 43,783 lines of code. Table 1 shows the breakdown of different components.

**AST parser.** We design a general AST parser to handle common features of different DBMSs, and customize the parser for each DBMS to support implementation-specific features. Our implementation is based on Bison 3.3.2 and Flex 2.6.4. The grammar of our AST parser conforms to the specification described in the official DBMS documents. We support most grammars in the specification, but leave alone some parts that are related to administrative functionalities. In this way, we can focus on testing those grammars related to database manipulation.

**Fuzzer.** We build Squirrel on top of AFL 2.56b, and replace its mutator with our syntax-preserving mutator and semantics-guided instantiator. When the fuzzer finds an interesting test case, we save its stripped IRs into the IR library. We drop the database after each query to minimize the interplay between different queries.

**Effort of Adoption.** The effort of adopting Squirrel to other DBMSs could be DBMS-dependent. First, we should customize the general parser to support the unique features of the target DBMS. Second, we will write semantic relation rules according to the grammar. Third, if the DBMS runs in client-server mode (e.g., MySQL and PostgreSQL), we need to implement a client for it. In our cases, it took one of our author one day to customize the parser, and less than six hours to implement the semantic rules and the client for each DBMS we tested. We believe it should take no more than two days to adopt Squirrel to another DBMS.

## 8 Evaluation

We applied Squirrel on real-world relational DBMSs to understand its effectiveness on finding memory error bugs. Specifically, our evaluation aims to answer the following questions:
- Can Squirrel detect memory errors from real-world production-level DBMSs? (§8.1)

Table 2: Compatibility between fuzzers and DBMSs. MySQL only permits C/S mode, which is not supported by the last three fuzzers. SQLsmith does not support MySQL's grammar. QSYM supports fuzzing PostgreSQL in the single mode, GRIMOIRE cannot compile it, while Angora cannot run it.

| | Squirrel | AFL | SQLsmith | QSYM | Angora | GRIMOIRE |
|---|---|---|---|---|---|---|
| SQLite | ✔ | ✔ | ✔ | ✔ | ✔ | ✔ |
| PostgreSQL | ✔ | ✔ | ✔ | ✔ (single) | ✘ (execute) | ✘ (compile) |
| MySQL | ✔ | ✔ | ✘ (interface) | ✘ (C/S) | ✘ (C/S) | ✘ (C/S) |

- Can Squirrel outperform state-of-the-art testing tools? (§8.2)
- What are the contributions of language correctness and coverage-based feedback in DBMS testing? (§8.3)

**Benchmarks.** We select three widely used DBMSs for extensive evaluation, including SQLite [4], PostgreSQL [3], MySQL [2]. We also test MariaDB [1] with Squirrel just for finding bugs. We compile them with default configurations and compilation options. We compare Squirrel with five fuzzers, including the mutation-based fuzzers AFL [66] and Angora [26], the hybrid fuzzer QSYM [65], the structural fuzzer GRIMOIRE [21] and the generation-based fuzzer SQLsmith [5]. We try to run as many tests as possible, but as shown in Table 2, we encounter several compatibility issues. As MySQL requires a client to send the query (i.e., C/S mode), QSYM, Angora and GRIMOIRE cannot test it directly. SQLsmith does not officially support MySQL due to the lack of interfaces [14]. PostgreSQL supports both C/S mode and single mode, and we can use QSYM to test it in the single mode. However, GRIMOIRE cannot compile PostgreSQL successfully to a single static binary; Angora can compile it but cannot run the binary. We are actively seeking potential solutions.

**Seed Corpus.** We collect seed inputs from the official Github repository of each DBMS, where the unit tests usually cover most types of queries. We feed the same seeds to all the six fuzzers in our evaluation except SQLsmith which does not need any initial inputs.

**Setup.** We perform our evaluation in a Ubuntu 16.04 system, on a machine that has Intel Xeon CPU E5-2690 (2.90GHz) with 16 cores and 188GB RAM. We use afl-clang with llvm mode to instrument tested DBMSs, and adopt edge-coverage for the feedback. Considering the large codebase of DBMSs, we use a bitmap with 256K bytes to mitigate the path collision issue [31]. Angora uses a 1024K-byte bitmap, the default size by design. For bug detection, due to the time limit and the implementation progress, we have run Squirrel to test SQLite for 40 days, MySQL and PostgreSQL for 11 days, and MariaDB for 1 day. For other evaluations, we run each fuzzing instance (fuzzer+DBMS) for 24 hours and repeat the process five times. Each fuzzing instance runs separately in a docker with one CPU and 10G memory. We report the average results to reduce the random noise and provide the p-values in Table 6 of Appendix.

### 8.1 DBMS Bugs

Squirrel has successfully detected 63 bugs from tested DBMSs, including 51 bugs from SQLite, 7 from MySQL, and 5 from MariaDB. Table 3 shows the details of the identified bugs. We have responsibly reported all the bugs to corresponding DBMS developers and have received their positive feedback. At the time of paper writing, 52 of

**Table 3: Detected bugs.** SQUIRREL found 63 bugs, including 51 from SQLite, 7 from MySQL and 5 from MariaDB. SQLite 3.31 was under development and we tested the latest version on Github. †: SQLite does not has severity score for bugs. ? means the severity has not been decided by developers.

**UAF**: use-after-free. **BOF**: buffer overflow of Global (**G**), Heap (**H**), and Stack (**S**).
**BUF**: buffer underflow. **AF**: assertion failure. **OOM**: out of memory. **UB**: undefined behavior.

| ID | Type | Function | Status | Severity† | Reference |
|----|------|----------|--------|-----------|-----------|
| **SQLite v3.30.1, 300K LoC** | | | | | |
| 1 | BOF | PRAGMA integrity_check | Fixed | Critical | CVE-2019-19646 |
| 2 | NP | lookupName | Fixed | Critical | CVE-2019-19317 |
| 3 | UAF | WITH | Fixed | High | CVE-2019-20218 |
| 4 | BOF | exprListAppendList | Fixed | High | CVE-2019-19880 |
| 5 | BOF | ZipFile extension | Fixed | High | CVE-2019-19959 |
| 6 | NP | zipfileUpdate | Fixed | High | CVE-2019-19925 |
| 7 | NP | parser | Fixed | High | CVE-2019-19926 |
| 8 | NP | LEFT JOIN optimization | Fixed | High | CVE-2019-19923 |
| 9 | SBOF | ALTER TABLE | Fixed | Medium | CVE-2019-19645 |
| 10 | NP | JOIN INDEX | Fixed | Medium | CVE-2019-19242 |
| 11 | NP | parser | Fixed | Medium | CVE-2019-19924 |
| 12 | BOF | propagateConstantExprRewrite | Fixed | Medium | CVE-2020-6405 |
| 13 | UB | fopen/fopen64 | Fixed | - | 0c4f820 |
| 14 | GBOF | sqlite3VdbeMemPrettyPrint | Fixed | - | 5ca0632 |
| 15 | AF | sqlite3GenerateConstraintChecks | Fixed | - | ad5f157 |
| 16 | AF | IN expression optimization | Fixed | - | b97f353 |
| 17 | AF | whereLoopAddOr | Fixed | - | 9a1f2e4 |
| 18 | AF | WHERE with OR opt. | Fixed | - | a4b2df5 |
| 19 | AF | wherePathSatisfiesOrderBy | Fixed | - | 77c9b3c |
| 20 | AF | Bytecode OP_DeferredSeek | Fixed | - | be3da24 |
| 21 | AF | WHERE | Fixed | - | 4adb1d0 |
| 22 | AF | WHERE flag setting | Fixed | - | 118efd1 |
| 23 | AF | Bytecode OP_ResultRow release | Fixed | - | 02ff747 |
| 24 | AF | sqlite3SelectReset | Fixed | - | aa328b6 |
| 25 | AF | Bytecode OP_SCopy | Fixed | - | 629b88c |
| 26 | AF | scalar subquery | Fixed | - | 629b88c |
| 27 | AF | Bytecode OP_ResultRow | Fixed | - | 02ff747 |
| 28 | AF | SELECT | Fixed | - | fbb6e9f |
| 29 | AF | WHERE | Fixed | - | f1bb31e |
| 30 | AF | PRAGMA encoding | Fixed | - | b5f0e40 |
| **SQLite v3.31 (under development), 304K LoC** | | | | | |
| 31 | GBOF | ZipFile extension | Fixed | - | 8d7f44c |
| 32 | HBOF | ZipFile extension | Fixed | - | a194d31 |
| 33 | HBUF | ZipFile extension | Fixed | - | 8d7f44c |
| 34 | UAF | sqlite3GenerateConstraintChecks | Fixed | - | 6d67aff |
| 35 | NP | VTable | Fixed | - | c7a5ff4 |
| 36 | NP | ORDER BY Windows Function | Fixed | - | 73bacb7 |
| 37 | NP | SF_Aggregate flag setting | Fixed | - | 9e10f9a |
| 38 | NP | USING | Fixed | - | 0824d5b |
| 39 | NP | ZipFile extension | Fixed | - | 0d21eae |
| 40 | NP | LEFT JOIN uses values from IN | Fixed | - | 74ebaad |
| 41 | AF | WHERE | Fixed | - | b592d47 |
| 42 | AF | NEVER marco can be true | Fixed | - | 78b5220 |
| 43 | AF | impliesNotNullRow | Fixed | - | aef8167 |
| 44 | AF | Code Generator for inline function | Fixed | - | 25c4296 |
| 45 | AF | scalar SELECT w/ WINDOW | Fixed | - | 4ea562e |
| 46 | AF | Code Generator for sub query | Fixed | - | fc705da |
| 47 | AF | AND optimization | Fixed | - | 2b6e670 |
| 48 | AF | Bytecode OP_Move | Fixed | - | 4cbd847 |
| 49 | AF | Bytecode OP_Copy-coalesce opt. | Fixed | - | 9099688 |
| 50 | AF | sqlite3ExprCodeIN | Fixed | - | f6ea97e |
| 51 | AF | whereTermPrint | Fixed | - | 6411d65 |
| **MySQL v8.0, 4250K LoC** | | | | | |
| 52 | OOM | WITH optimization | Verified | Critical | ID98190 |
| 53 | NP | JOIN optimization | Fixed | Serious | ID98119 |
| 54 | NP | JOIN optimization | Verified | ? | ID99438 |
| 55 | NP | UPDATE optimization | Verified | ? | ID99424 |
| 56 | AF | SELECT | Verified | ? | ID99420 |
| 57 | AF | INDEX | Verified | ? | ID99421 |
| 58 | AF | CREATE TABLE | Verified | ? | ID99454 |
| **MariaDB v10.5.3, 3641K LoC** | | | | | |
| 59 | BOF | UPDATE | Verified | ? | MDEV22464 |
| 60 | BOF | UPDATE | Verified | ? | MDEV22476 |
| 61 | AF | JOIN | Verified | ? | MDEV22461 |
| 62 | AF | SELECT | Verified | ? | MDEV22462 |
| 63 | AF | Array OOB | Verified | ? | MDEV22463 |

all the bugs have been fixed. 12 of them got CVE numbers due to the severe security consequences. As a comparison, the OSS-Fuzzer project launched by Google [34] have extensively tested the first thee DBMSs, and found 19 bugs from SQLite in three years, 15 bugs from MySQL in four months and no bugs from PostgreSQL. We inspect the MySQL bugs detected by OSS-Fuzzer and find that all of them happen at the very beginning of MySQL's logic: before the parsing phase. The proof-of-concepts (PoC) are not even valid SQL queries but just some random bits. Therefore, we believe our fuzzer can find bugs from DBMSs more effectively. We plan to integrate our tool into OSS-Fuzzer to improve the security of DBMSs.

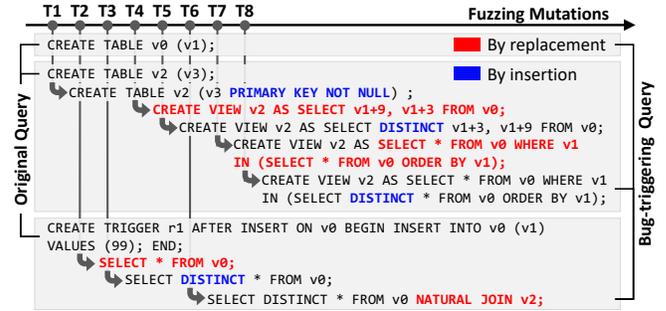

**Figure 8: Mutations to trigger the 11-year old bug.** SQUIRREL takes eight steps, including four insertions and four replacements to produce the bug-triggering query from the original one.

**Bug Diversity.** The 63 bugs in Table 3 cover almost all common types of memory errors, showing that SQUIRREL can improve DBMS security from a variety of aspects. In particular, buffer overflows and use-after-free bugs are commonly believed to be exploitable, whereas SQUIRREL found **12** bugs and **2** bugs, respectively. SQUIRREL also detected **33** assertion failures from SQLite, which indicate that the executions reach unexpected states. Even worse, assertion checks are disabled in the released binary, which may lead to severe security problems. For example, in case study 3, an assertion failure results in a severe use-after-free vulnerability.

**Case Study 1: An 11-Year-Old Bug.** SQUIRREL detected a bug introduced to SQLite 11 years ago (ID 16 in Table 3, PoC in Listing 1 of Appendix), which lies in an optimization routine of the IN clause. Specifically, isCandidateForInOpt checks various conditions to determine whether the subquery inside the IN clause can be optimized or not. One of these checks should make sure the subquery does not have any GROUP BY clause. As the SQL grammar does not allow GROUP BY in an IN clause, the developers "were unable to find a test case for"[1] this condition, and thus converted the check "into an assert() by check-in [...] (2009-05-28)". The assertion is disabled in the released version of SQLite. SQUIRREL found that if two queries with DISTINCT are joined by NATURAL JOIN, SQLite will *internally* set the GROUP BY property to these queries. When such queries are used in an IN clause, it will make the previous assertion fail. However, the released SQLite will continue the optimization incorrectly, and may lead to unexpected consequences, like wrong results.

SQUIRREL found this 11-year-old bug within 14 minutes through only eight mutations. Figure 8 shows the eight steps for SQUIRREL to generate the bug-triggering query from a benign one. We denote $T_n$ as the $n$th mutation. The original query contains three CREATE statements: the first CREATE does not have to change; the second CREATE is changed five times with three insertions ($T_1$, $T_5$ and $T_8$) and two replacements ($T_4$ and $T_7$); the last CREATE is changed three times with two replacements ($T_2$ and $T_6$) and one insertion ($T_3$). Each round of mutation provides a new syntactical structure, and keeps both syntactical and semantic correctness. The final query satisfies the conditions for failing the assertion, as SQLite will put the last SELECT into the IN of the second statement, which makes the subquery of IN contain two naturally joined SELECTs.

---
[1]Quote from the message of the fix commit.

**Case Study 2: Database Leakage.** Squirrel identified a heap-based buffer overread vulnerability (ID 5 in Table 3, PoC in Listing 2 of Appendix), which allows attackers to read arbitrary data in the memory space. This bug entitles attackers to potentially access all databases stored in the SQLite DBMS. As SQLite is widely used as a multi-user service, attackers can retrieve the data of other users, which by default they have no access. Even if the database is explicitly deleted by its owner, attackers can still steal it from its memory residue. Other than stealing databases, this bug also enables attackers to read sensitive-critical information that may allow attackers to build further attacks, like remote code execution. For example, reading the code page address will help attackers bypass randomization-based defenses, while leaking the stack canary will make stack buffer-overflow exploitable again.

**Case Study 3: Assertion Failure Leading to Use-After-Free.** We inspected several assertion failures and found one (ID 3 in Table 3, PoC in Listing 3 of Appendix) finally leads to a **high-severity** use-after-free bug (score 7.5/10). An assertion affirms that a predicate is always true whenever the execution reaches the assertion point. Otherwise, the developer considers the program running into an unexpected state. This bug makes SQLite fail a pParse->pWith assertion, as pWith is a dangling pointer due to a failed creation of a circular view. In the debug mode, SQLite will terminate the execution after the assertion failure. However, the released binary disables all assertions. Given the bug-triggering input, SQLite will keep running in the unexpected state and finally trigger the use-after-free bug.

**Case Study 4: Fuzzing as Regression Test.** Squirrel can effectively find newly-introduced bugs, and thus can be used for rapid regression test. For example, the bug with ID 35 in Table 3 (PoC in Listing 4 of Appendix) only exists for less than one day before we found it. The bug with ID 38 (PoC in Listing 5 of Appendix) got detected, reported and even fixed in just one hour after its existence. The commit introducing this bug was intended to fix another issue related to the generated column functionality. However, the fix was not completely correct and thus introduced a new problem in the USING clause. These two cases show that Squirrel can do fast and effective regression testing for DBMSs.

## 8.2 Comparison with Existing Tools

We compare Squirrel with five state-of-the-art fuzzers in different aspects to understand its strength and weakness in testing DBMSs. Figure 9 shows our evaluation results, including the number of unique crashes, the number of unique bugs, the number of new edges, the syntax correctness and the semantic correctness. The p-values of our evaluations are shown in Table 6 in the Appendix. Most of the p-values are less than 0.05, which means the differences between Squirrel's results and others' are statistically significant. We will discuss exceptional high p-values case by case.

**Unique Crashes.** We utilize the edge-coverage map to calculate the number of unique crashes, and show the results of fuzzing SQLite in Figure 9(a). We exclude the results of PostgreSQL and MySQL as only Squirrel finds few crashes in MySQL, and none of the other fuzzing instances find any crashes within 24 hours. Squirrel detects the first crash of SQLite within four minutes, and totally detects about 600 unique ones. AFL catches the first crash in 32 minutes and gets 30 unique ones in total, while QSYM discovers the first one in 14 minutes and finally collects 13 crashes. Angora, GRIMOIRE and SQLsmith cannot find any crashes.

**Table 4: Distribution of SQLite bugs found by fuzzers.** We perform the evaluation for 24 hours, repeat for 5 times and report the average results. We update SQLite binary per-hour to fix identified bugs. † ID in Table 3.

| Target | ID† | Type | Squirrel | AFL | SQLsmith | QSYM | Angora | GRIMOIRE | !semantic | !feedback |
|---|---|---|---|---|---|---|---|---|---|---|
| SQLite | 2 | NULL Ptr Deref | ✔ | ✔ | ✗ | ✔ | ✗ | ✗ | ✔ | ✔ |
| SQLite | 3 | Use-After-Free | ✔ | ✗ | ✗ | ✗ | ✗ | ✗ | ✗ | ✗ |
| SQLite | 4 | Buffer Overflow | ✔ | ✗ | ✗ | ✗ | ✗ | ✗ | ✗ | ✗ |
| SQLite | 5 | Buffer Overflow | ✔ | ✗ | ✗ | ✗ | ✗ | ✗ | ✗ | ✗ |
| SQLite | 6 | NULL Ptr Deref | ✔ | ✗ | ✗ | ✗ | ✗ | ✗ | ✗ | ✗ |
| SQLite | 9 | Stack Overflow | ✔ | ✗ | ✗ | ✗ | ✗ | ✗ | ✗ | ✗ |
| SQLite | 10 | NULL Ptr Deref | ✔ | ✗ | ✗ | ✗ | ✗ | ✗ | ✗ | ✗ |
| SQLite | 11 | NULL Ptr Deref | ✔ | ✗ | ✗ | ✗ | ✗ | ✗ | ✗ | ✗ |
| SQLite | 14 | Buffer Overflow | ✔ | ✗ | ✗ | ✗ | ✗ | ✗ | ✗ | ✗ |

As we can see, recent advanced fuzzing tools do not significantly outperform AFL. In some cases, they may even find fewer unique crashes. We believe such result is reasonable due to the indeterministic characteristic of fuzzing and the strict semantic requirements for testing DBMS systems.

**Unique Bugs.** We map each crash to the corresponding bug based on the official patches. In SQLite, the 600 crashes found by Squirrel only belong to two bugs, while the 30 crashes detected by AFL and the 13 found by QSYM belong to the same one bug. Since the small number of bugs is not statistically useful, we take a different strategy to get more bugs: after every hour, we check the detected crashes (if any), map them to real bugs and patch them to avoid future similar crashes. We show the result of the new strategy in Figure 9(b). This method is effective for Squirrel to find more bugs (from two to nine), as after each patching, Squirrel can almost immediately find one more new bug. AFL and QSYM can only find one bug within one hour, and have no progress even with the patching. Table 4 shows the distribution of detected bugs, where the only bug found by AFL and QSYM is also covered by Squirrel.

**New Edges.** Squirrel identifies 2.0×-10.9× *more* new edges than mutation-based tools, and achieves a comparable result to the generation-based tester SQLsmith. Figure 9(c), (f) and (i) show the increase of new edges of SQLite (S), PostgreSQL (P) and MySQL (M), respectively. Squirrel outperforms other fuzzers in eight comparisons out of nine: it collects 6.6× (S), 4.4× (P) and 2.0× (M) more new edges than AFL, 7.7× (S) more than SQLsmith, 3.6× (S) and 10.9× (P) more than QSYM, 2.3× (S) more than Angora, and 3.3× (S) more than GRIMOIRE. The only exception comes from fuzzing PostgreSQL with SQLsmith, where Squirrel collects 89.3% new edges of that by SQLsmith. This is not a surprise considering that SQLsmith is designed to handle the specific grammar of PostgreSQL. Since SQLsmith performs slightly better than Squirrel on PostgreSQL, the related p-value in Table 6 is larger than 0.05.

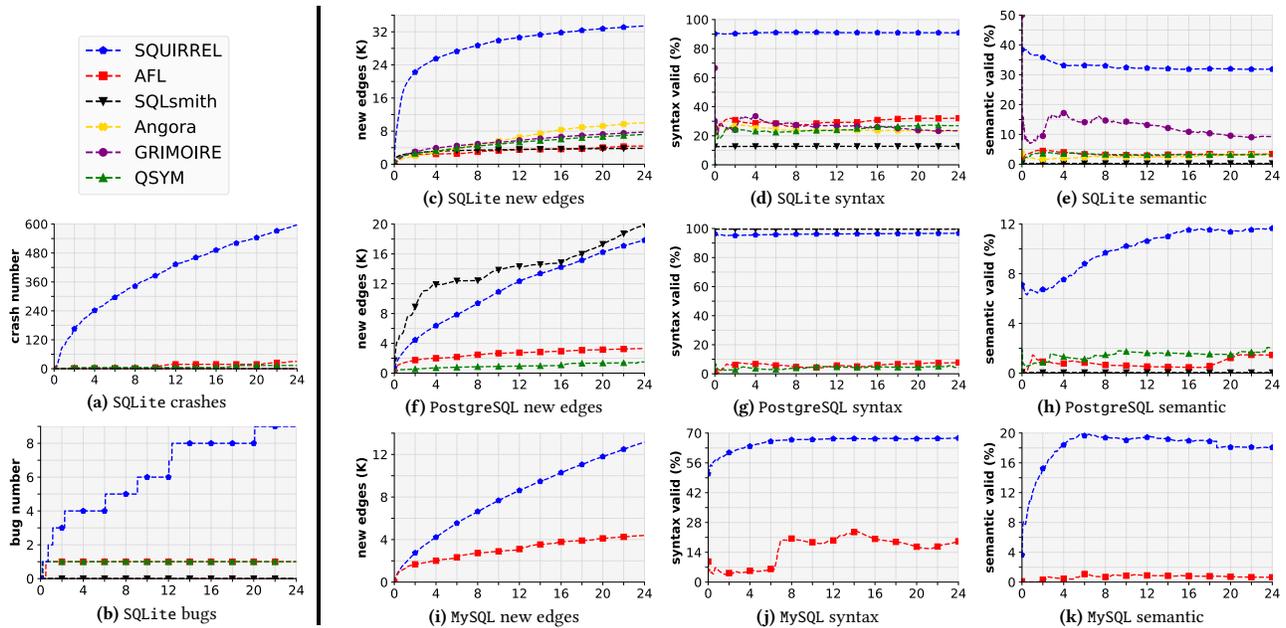

**Figure 9: Comparison with existing tools.** Figures (a) and (b) show the numbers of unique crashes and the numbers of unique bugs for fuzzing SQLite. Figures (c)-(k) show the number of new edges, the syntax correctness and the semantic correctness for each fuzzing instance. We run each fuzzing instance for 24 hours, repeat each fuzzing for five times and report the average results.

**Syntax Validity.** SQUIRREL achieves 1.8×-20.9× *higher* syntax correctness than mutation-based tools, and gets a comparable result to SQLsmith. Figure 9(d), (g) and (j) show the change of syntactic validity during testing SQLite (S), PostgreSQL (P) and MySQL (M), respectively. SQUIRREL achieves 1.8× (S), 11.5× (P) and 2.5× (M) higher syntax correctness than AFL, 6.1× (S) higher than SQLsmith, 2.4× (S) and 20.9× (P) higher than QSYM, 2.8× (S) higher than Angora, and 2.9× (S) higher than GRIMOIRE. The exception comes from fuzzing PostgreSQL with SQLsmith, where SQUIRREL achieves 97.1% syntactic validity of that by SQLsmith. Again, we believe the reason is that SQLsmith is highly customized for the specific grammar of PostgreSQL. For example, SQLsmith only achieves 12.7% syntax correctness in fuzzing SQLite, while gets almost 100% syntax accuracy when fuzzing PostgreSQL. The p-value of SQUIRREL vs SQLsmith on PostgreSQL is more than 0.05 due to the similar results.

**Semantic Validity.** SQUIRREL achieves 2.4×-243.9× *higher* semantic correctness than other tools. Figure 9(e), (h) and (k) show the trend of semantic validity during testing SQLite (S), PostgreSQL (P) and MySQL (M), respectively. SQUIRREL achieves 8.3× (S), 7.0× (P) and 27.0× (M) higher semantic correctness than AFL, 125.4× (S) and 243.9× (P) higher than SQLsmith, 8.8× (S) and 4.7× (P) higher than QSYM, 8.3× (S) higher than Angora, and 2.4× (S) higher than GRIMOIRE. Interestingly, although SQLsmith performs slightly better on testing PostgreSQL with respect to new edges and syntax correctness, SQUIRREL achieves significantly higher accuracy on semantics. Another worth-noting observation is that AFL actually generates more correct inputs for PostgreSQL than SQUIRREL (2.2×, see Table 7 in Appendix), but still achieves lower edge coverage. This indicates that larger numbers of correct queries or higher correct rate cannot guarantee to explore more program states. One extreme example is to keep using the same correct query, which will have more executions (no generation overhead) and 100% correctness. But apparently, it will lead to no increase in code coverage. The strength of SQUIRREL stems from both the syntax-preserved mutation, which generates queries of various structures, and the semantics-guided instantiation, which infers semantic relationships between arguments to assist query synthesis.

> Overall, SQUIRREL outperforms all mutation-based tools, even if they are augmented with taint analysis or symbolic execution, or take structural information into consideration. It achieves comparable results to SQLsmith, which is customized for PostgreSQL. More importantly, SQUIRREL can detect significantly more bugs than all other tested tools.

### 8.3 Contributions of Validity and Feedback

To understand the contribution of different factors in SQUIRREL, specifically, syntax-preserving mutation, semantic-guided instantiation and coverage-based feedback, we perform unit tests by disabling each factor and measure various aspects of the fuzzing process. The results are given in Figure 10. In SQUIRREL$_{!semantic}$, we only disable the semantics-guided instantiation; in SQUIRREL$_{!feedback}$, we only disable the coverage-based feedback; SQUIRREL$_{!semantic}^{!syntax}$ disables both the semantics-guided instantiation and the syntax-correct mutation, and is in fact the same as AFL. Since our semantics-guided instantiation requires syntax-correct queries, we cannot create a version that only disables the mutation. We also exclude the all-disabled setting, which will be the dumb mode of AFL.

The p-values of our evaluations are shown in Table 6. Most p-values are less than 0.05, showing that the differences between

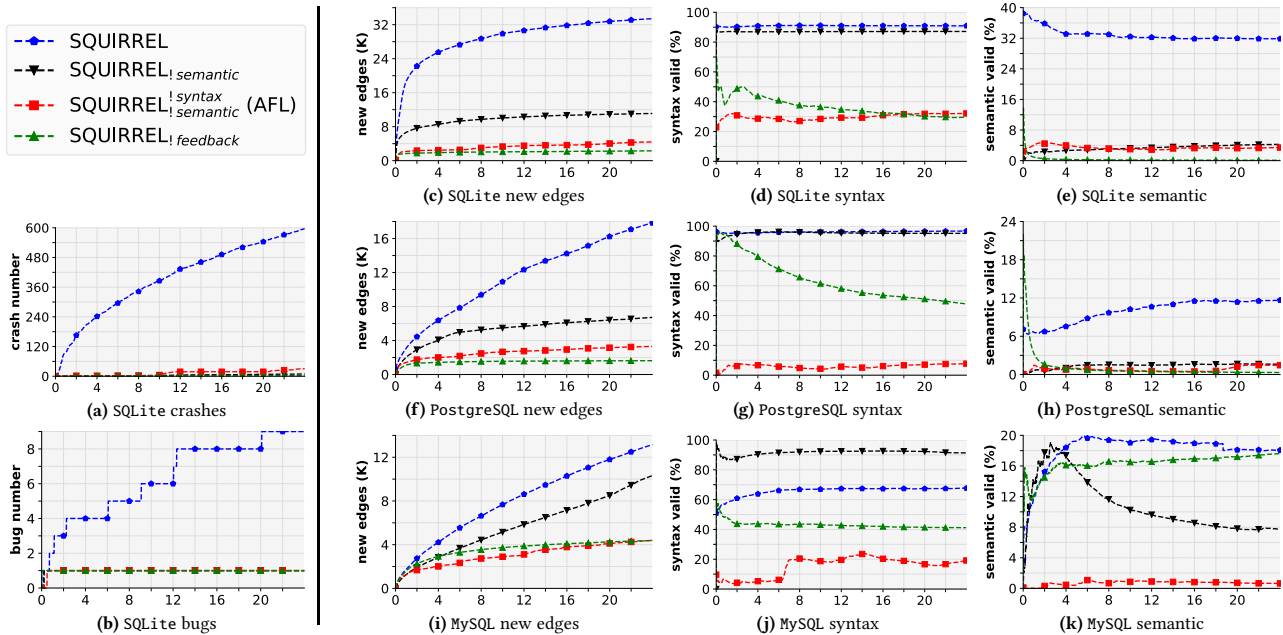

**Figure 10: Contributions of Validity and Feedback.** Figures (a) and (b) show the numbers of unique crashes and the numbers of unique bugs for fuzzing SQLite. Figures (c)-(k) show the number of new edges, the syntax correctness and the semantic correctness for each fuzzing instance. We run each fuzzing instance for 24 hours, repeat each fuzzing for five times and report the average results.

Squirrel's results and others' are statistically significant. We will explain exceptional p-values higher than 0.05.

**Unique Crashes.** Figure 10(a) shows the number of unique crashes found in SQLite by each setting. Similarly, we skip the results of PostgreSQL and MySQL due to the small number of crashes during the 24-hour evaluation. The full-featured Squirrel achieves the best results. First, Squirrel finds the first crash within four minutes. Squirrel$_{!semantic}$ takes 60× more time to detect the first crash (261 minutes). Interestingly, Squirrel$_{!semantic}^{!syntax}$ finds the first crash within 32 minutes — worse than Squirrel, but better than Squirrel$_{!semantic}$. Considering that the latter runs 220 queries per second while the former can execute 507 (i.e., 1.3× faster), we believe the advantage of Squirrel$_{!semantic}^{!syntax}$ over Squirrel$_{!semantic}$ mainly comes from the faster generation speed. Squirrel$_{!feedback}$ takes 700 minutes to find the first crash. The total number of unique crashes follows the same pattern, where Squirrel, Squirrel$_{!semantic}$, Squirrel$_{!semantic}^{!syntax}$ and Squirrel$_{!feedback}$ detect 600, 30, 10 and 3 crashes, respectively. The results above show that all three factors of Squirrel contribute critically to the crash detection. Further, coverage-based feedback plays the most important role, while syntax-only testing cannot beat AFL.

**Unique Bugs.** We take the same strategy to measure the unique bugs, where we patch the detected bug after each hour. Figure 10(b) shows the results. The full-featured Squirrel finds nine unique bugs, while other variants only detect one unique bug. As shown in Table 4, the only bug found by Squirrel variants (!feedback and !semantic) is also covered by the full-featured version.

**New Edges.** Figure 10(c), (f) and (i) demonstrate an (almost) consistent pattern of finding new edges for SQLite, PostgreSQL and MySQL: Squirrel > Squirrel$_{!semantic}$ > Squirrel$_{!semantic}^{!syntax}$ > Squirrel$_{!feedback}$. The coverage-based feedback helps achieve 2.0× more new edges for fuzzing SQLite, PostgreSQL and MySQL. Syntax-correctness helps find 1.0×-1.5× more edges than AFL, while semantic correctness further improves the number by 0.3×-1.7×. This result shows that improving the syntax-correctness or semantic correctness helps reach more DBMS states.

**Syntax Validity and Semantic Validity.** Figure 10(d), (g) and (j) show the syntax changes during the tests of three DBMSs, while Figure 10(e), (h) and (k) show the semantic changes. In most cases, Squirrel achieves the best validity while AFL reaches the worst. This result is reasonable as we design Squirrel to get better language validity while AFL randomly mutates SQL queries. However, we can find some interesting anomalies from the figures. First, Squirrel$_{!semantic}$ achieves similar syntax accuracy as Squirrel in Figure 10(d) and (g), which shows that improving semantic correctness will not increase syntax correctness. In fact, the instantiation may decrease the syntax correctness, like in Figure 10(j), as it tends to remove short queries. Short queries are more likely syntactically correct, but our instantiator cannot fix their semantics. For example, SELECT a FROM b is semantically incorrect as no table b exists. As SQLsmith performs similarly or even better than Squirrel, the p-values for PostgreSQL and MySQL are larger than 0.05.

Second, in Figure 10(j) Squirrel$_{!feedback}$ has the similar semantic correctness as Squirrel. This result seems to suggest that feedback has no impact on MySQL's semantic correctness. However, further inspection reveals that Squirrel$_{!feedback}$ produces very divergent semantic correctness: among five experiments, two achieve high correctness over 40%, while the other three have a low rate below 10%. We find that the initial seeds in MySQL are smaller than those in `SQLite` and `PostgreSQL`. These small seeds might induce SQUIRREL$_{!feedback}$ to keep producing correct but simple and repeated inputs for `MySQL`. Due to the randomness of `MySQL` results, the p-values in Table 6 of Appendix are larger than 0.05. However, Figure 10(i) shows that the queries generated by SQUIRREL are more diverse in structures than those of SQUIRREL$_{!feedback}$, as SQUIRREL finds much more execution paths with similar semantic correctness.

> Overall, syntax, semantics and feedback play critical roles in SQUIRREL to find more memory errors from DBMSs. The coverage-based feedback has the most impact, while syntactic correctness and semantic correctness have mixed influences. The final result is an interplay between all three factors.

## 9 Discussion

We discuss several limitations of our current implementation of SQUIRREL and our plan to address them in future work.

**DBMS-Specific Logic.** Although our design of SQUIRREL is DBMS-agnostic, we find that incorporating program-specific features always helps achieve a better testing result. First, each DBMS implements a dialect of SQL, which is either almost the same as the official one, like `SQLite` [4], or significantly different in many features, like `PostgreSQL` [3]. SQUIRREL fully supports the general grammar of SQL, and also contains patches for different dialects. Due to this reason, SQUIRREL works well on `SQLite` (51 bugs), but only triggers few bugs in `PostgreSQL`, `MySQL` and `MariaDB`. We plan to implement more accurate grammar of different SQL dialects to improve our fuzzing efficacy. Second, DBMSs may adopt extra checks before executing the query. For example, we find that `PostgreSQL` requires type correctness between all operands, and comparisons between integers and floating points are not allowed. `SQLite` does not check anything but will automatically perform type-casting during the execution, while `MySQL` only gives a warning for mismatched types. We plan to implement the type consistency relation in our semantics-guided instantiation for testing `PostgreSQL`.

**Relation-Rule Construction.** SQUIRREL relies on relation rules to infer data dependencies between different operands. Currently, we write the relation rules based on our domain knowledge. We have two of our authors spend two hours to write these rules, which merely covers 133 clauses. To alleviate developers from this tedious process, we plan to adopt techniques to automatically infer these rules. For example, with data-flow analysis, we can figure out the expected relationship between each operand. Alternatively, we can try machine learning techniques to automatically capture the relationship from a large number of normal executions.

**Collisions in Code-Coverage.** SQUIRREL relies on the feedback mechanism of AFL to guide the query selection, which unfortunately suffers from the collision issue [31]. By default, AFL uses a bitmap with 64K entries to record the branch coverage, one for each branch. For small programs with few branches, this method works very well. However, DBMSs contain tens of thousands of branches and thus testing DBMSs with AFL has a severe collision problem. For example, `SQLite` has about 20,000 distinct branches, and 14% of them share the bitmap entries with others [67]. During our evaluation, we enlarge the bitmap to 256K to mitigate the collision issue. In the next step, we plan to adopt the solution proposed in the CollAFL to eliminate the collision problem [31].

**Alternative Feedback Mechanisms.** Recent software testing practice widely adopts code coverage to guide mutation-based fuzzing [22, 33, 43, 54, 66]. However, during our evaluation, we find potential harmful code coverage that hinders the generation of semantics-correct queries. Especially, at the beginning of testing, grammar-incorrect queries trigger many new branches in the fault-handling code. The coverage-based feedback guides SQUIRREL to focus on these inputs instead of the original semantics-correct queries. Recent works for fuzzing language compilers and interpreters mention a similar observation [35, 62]. We plan to investigate this problem and develop solutions to mitigate it, like dropping inputs that trigger new branches in short executions.

## 10 Related Work

**Detecting Logic and Performance Bugs in DBMSs.** DBMSs have been heavily tested for logic and performance defects [53, 56, 61, 64]. RAGS detects correctness bugs in DBMS through differential testing [56]. It generates and executes queries in multiple DBMSs. Any inconsistency among results indicates at least one DBMS contains bugs. SQLancer constructs queries to fetch a randomly selected row from a table [53]. The tested DBMS may contain a bug if it fails to fetch the row. QTune [41] is a database tuning system based on a deep reinforcement learning model, which can efficiently tune the database configurations for the best performance. Apollo uses differential testing to find performance bugs [37]. It generates and runs queries in two versions of the same DBMS. If two executions take significantly different times, the query triggers a performance regression bug. BmPad runs predefined test suites in the target DBMS, and reports performance bugs if the execution time exceeds the threshold [52]. SQUIRREL differs from these works by focusing on detecting memory corruption bugs, which can cause severe security consequences.

**Generation-based DBMS Testing.** Generation-based testing is commonly used to test DBMSs [5, 47, 59]. It can generate syntax-correct test cases efficiently but seldom guarantees the semantic correctness. QAGen shows that ensuring perfect semantic correctness is an NP-complete problem [44]. Instead, it provides an approximate solution to improve the semantic correctness. Several works reduce the generation to the SAT problem [15, 45] and use an SAT solver (*e.g.*, Alloy [17]) to provide potential solutions. Generation-based fuzzers usually require the schema of some initial databases for query generation. Bikash Chandra *et al.* [25] proposed a way to generate initial databases that can cover most types of SQL queries. SQLsmith is the state-of-the-art generation-based DBMS tester [5]. It collects the schemas from the initial databases and generates limited types of queries, like SELECT, to ensure that the database is unchanged, which restricts the code coverage. In contrast, SQUIRREL generates context-free test cases and does not rely on specific databases or schemas. It starts with an empty database and creates proper content before using them for testing.

**Mutation-based DBMS Testing.** Recently, mutation-based fuzzers [19, 23, 26, 27, 32, 40, 43, 57, 65, 66] have gained great success in finding memory errors. However, they are implemented as general fuzzers and are unaware of the structure of inputs. Though some of them adopt advanced techniques such as taint analysis or symbolic execution [19, 26, 65], they still cannot deeply test programs like DBMSs that accept highly structural inputs with correct semantics. Tim Blazytko *et al.* [21] proposed a way to utilize grammar-like combinations to synthesize highly structured inputs, but most of its generated test cases in SQL are still syntactically incorrect. Hardik Bati *et al.* [20] proposed to mutate SQL statements by adding or removing grammar components. They can likely preserve the syntactical correctness, but cannot guarantee the semantic correctness. Recent works [18, 50] tend to improve semantic correctness of the generated inputs, but SQL has more restrict semantic requirements and none of them show their effectiveness in testing DBMSs. Therefore, most test cases generated by these fuzzers fail either the syntax check or semantic check and have no chance to trigger deep logics, like optimization or execution. Squirrel overcomes these shortcomings with syntax-preserving mutation and semantic-guided instantiation and manages to detect bugs behind the deep logic.

## 11 Conclusion

We have proposed and implemented Squirrel to fuzz database management systems to find memory-related bugs. Our system employs two novel techniques, syntax-preserving mutation and semantics-guided instantiation, to help generate correct SQL queries. We evaluated Squirrel on four popular DBMSs: SQLite, MySQL, MariaDB and PostgreSQL, and found 51 bugs in SQLite, 7 in MySQL and 5 in MariaDB. Squirrel achieves at least 3.4 times of improvement in semantic correctness than that of current mutation-based and generation-based fuzzers, and triggers up to 12 times of improvement in code coverage than current mutation-based fuzzers. The results show that Squirrel is effective and efficient in testing database management systems.

## Acknowledgment

We thank the anonymous reviewers for their helpful feedback. The work was supported in part by the National Science Foundation (NSF) under grant CNS-1652790, and the Office of Naval Research (ONR) under grants N00014-16-1-2912, N00014-16-1-2265, N00014-17-1-2894, N00014-17-1-2895 and N00014-18-1-2662. Any opinions, findings, conclusions or recommendations expressed in this material are those of the authors and do not necessarily reflect the views of NSF or ONR.

**Algorithm 3:** Dependency Graph Construction.

**Input**    : *irSet*: The set of IRs that needs to be instantiated,
              *relationSet*: The set of predefined relations,
**Output**: *depencyGraph*: The dependency DAG

1  **Procedure** BuildDepGraph(*irSet, relationSet*)
2     resultGraphMap ← new HashMap();
3     **for** *each ir in irSet* **do**
4        resultGraphMap[ir] ← new Vector();
5     **for** *each ir ∈ irSet* **do**
6        **for** *each relation in relationSet* **do**
7           **if** *ir.DataType == relation.SecondElement* **then**
8              matchType ← relation.FirstElement;
9              candidateSet ← all the IRs of matchType in irSet that positions before ir in the query;
10             mappedTarget ← choose one IR from candidateSet according to the properties of relation;
11             result[mappedTarget].insert(ir);
12    **return** resultGraphMap

**Table 5: Definition of our SQL intermediate representation (IR).** Since | is alo a symbol in the definition of *symbol*, we use comma instead.

| | | | |
|---|---|---|---|
| *program* | $p$ | ::= | $s \mid s\ p$ |
| *statement* | $s$ | ::= | $v = e;$ |
| *expression* | $e$ | ::= | $\beta_1 \diamond \beta_2 \mid \rho$ |
| *operator* | $\diamond$ | ::= | $\diamond_p \diamond_m \diamond_s$ |
| *oprand* | $\beta$ | ::= | $v \mid$ NULL |
| *prefix* | $\diamond_p$ | ::= | $k$ |
| *mid* | $\diamond_m$ | ::= | $k$ |
| *suffix* | $\diamond_s$ | ::= | $k$ |
| *variable* | $v$ | ::= | $v_0, v_1, \ldots$ |
| *literal* | $\rho$ | ::= | string \| number |
| *keyword* | $k$ | ::= | NULL \| SELECT \| CREATE \| FROM \| WHERE \| ... |

## A  IR Definition

Table 5 shows the formal definition of our IR. Each program is a sequence of assignment statements in the static single assignment (SSA) form. Each statement has a left part, which is a variable, and a right part, which is an expression. An expression is either a literal (constant string or number) or an operation with an operator and two operands. An operator can have an optional prefix, middle or suffix. Benefit from designing IR like this, the mutation can be performed in a uniform way. Further, it makes translating IR back to a SQL query easier.

**Algorithm 2:** Convert IR back to SQL query.

**Input**    : *ir*: The pointer of the root IR,
**Output**: *sqlQuery*: The string of the SQL query

1  **Procedure** IRToString(*ir*)
2     **if** *ir.Data is not NULL* **then**
3        **return** ir.data
4     ResultString ← new String;
5     **if** *ir.Op.Prefix is not NULL* **then**
6        ResultString ← ResultString + ir.Op.Prefix
7     **if** *ir.LeftOperand is not NULL* **then**
8        LeftOperandString ← IRToString(*ir.LeftOperand*);
9        ResultString ← ResultString + LeftOperandString
10    **if** *ir.Op.Mid is not NULL* **then**
11       ResultString ← ResultString + ir.Op.Mid
12    **if** *ir.RightOperand is not NULL* **then**
13       RightOperandString ← IRToString(*ir.RightOperand*);
14       ResultString ← ResultString + RightOperandString
15    **if** *ir.Op.Suffix is not NULL* **then**
16       ResultString ← ResultString + ir.Op.Suffix
17    **return** ResultString

## B  Query Translation

**From SQL to IRs.** Before translating an SQL query to IRs, we first parse the queries into an AST using a parser. Then we use the following deep-first search algorithm to translate an AST into a set of IRs:

(1) If an AST node has child nodes, recursively translate all of its child nodes first.
(2) If an AST node has less than two children, and each of the children is already translated into an IR in (1), we just need to allocate a new IR and assign each of the children to be its operands. All the keyword before the left operand will be set as the prefix of the operator; all the keyword after the left operand and before the right operand (if exist) will be set as the middle of the operator; all the keyword after the right operand will be set as the suffix of the operator. Also, we fill in the new IR with other information such as operator and type.
(3) If an AST node has more than 2 children, then we put all IRs of its children into a queue in the same order as grammar. Each time we pop two IRs out from the head of the queue and perform step 2 using these two IRs. In this way, we get a new IR. Specifically, we guarantee the first IR we pop will be the left operand of the new IR. If the queue becomes empty now, we finish translating this node. If not, this IR doesn't correspond to any node in the AST but only part of a node. In this situation, it has a special type called Unknown. we push this IR to the head of the queue and repeat step 3.

From the algorithm, we can see that one single AST node can be translated into several IRs. Some of these IRs are immediate results of the translation with type kUnknown. Since all IRs are of the same format, they can be mutated in a uniform way, including these immediate IRs. In this way, our IR has a finer granularity than AST.

**From IRs to SQL.** Converting IRs back to SQL queries is required for fuzzing databases as DBMSs only accept SQL queries as inputs. Such conversion is easy for our IRs and can be achieved through a depth-first search. Algorithm 2 shows how to convert an IR back to a SQL query. When we translate an AST node to IRs, we firstly try to translate children of this node. Then we use every 2 results to generate a new IR until all the IRs corresponding to children node had been processed. For this reason, we can know all the children must be translated before the current node translate. Besides, after using two IRs to generate a new IR, it will be put to the head of the queue. Next time, it will be pop from the queue as the first IR, and become a left operand. This means, when we try to convert an IR back to SQL query, we need to convert left operand before the right operand. Line 2-4 shows if an IR carries data, it will not have other fields, just return with the data. Line 5-22 shows we should convert from left to right and combine them. Line 23 returns with the result of the combination.

## C  Dependency Graph Construction

The graph construction algorithm is shown in Algorithm 3 and briefly described as follows: We first collect all data-carrying IRs whose data information has been stripped off and thus needs to be instantiated. We create a node $v$ in the dependency graph $G$ for each collected IR. For every relation pair, say $N$ depends on $M$. For each node $v_n$ in type $N$, We search in $G$ to find all nodes that have the same type as $M$. Based on the property of the relation, we choose a node of type $M$, say $v_m$, and build an edge from $v_m$ to $v_n$.

## D  PoCs for Case Study

Listing 1 shows the final PoC for the 11-year-old bug. Listing 2 shows the PoC where the last line will dump the content of database v2 even through it has been deleted. Listing 3 shows the PoC for the UAF bug that results from an assertion. Listing 4 shows the PoC for the bug that exists for only one day. Listing 5 shows the PoC for the bug that exists for only one hour.

```
01 |CREATE TABLE v0 (v1);
02 |CREATE VIEW v2 AS SELECT * FROM v0 WHERE v1
03 |    IN (SELECT DISTINCT* FROM v0 ORDER BY v1);
04 |SELECT DISTINCT * FROM v0 NATURAL JOIN v2;
```

**Listing 1: Case Study 1: 11-Year-Old Bug**

```
01 |CREATE TABLE v0 (v1 char);
02 |INSERT INTO v0 VALUES ('1');
03 |CREATE TABLE v2(v3 text);
04 |INSERT INTO v2 VALUES ("1"*147), ("2"*42), ("3"*37);
05 |DROP TABLE v2;
06 |INSERT INTO V0 SELECT ZIPFILE(v1, NULL) FROM v0;
07 |INSERT INTO V0 SELECT ZIPFILE(v1, NULL) FROM v0;
08 |INSERT INTO V0 SELECT ZIPFILE(v1, NULL) FROM v0;
09 |SELECT HEX(v1) FROM v0;
```

**Listing 2: Case Study 2: Database leakage**

```
01 |CREATE TABLE v0 (a);
02 |CREATE VIEW v2 (v3) AS WITH x1 AS (SELECT * FROM v2);
03 |SELECT v3 AS x, v3 AS y FROM v2;
```

**Listing 3: Case Study 3: UAF from assertion**

```
01 |CREATE TABLE v0 ( v1 INTEGER PRIMARY KEY ) ;
02 |INSERT INTO v0 ( v1 ) VALUES ( 0 )
03 |    ON CONFLICT DO NOTHING ;
04 |CREATE VIRTUAL TABLE v2
05 |    USING rtree(v5 UNIQUE ON CONFLICT ABORT, v4, v3);
06 |SELECT 'a' FROM v0
07 |    LEFT JOIN v2 ON v4 = 10 OR v5 = 10 ;
08 |SELECT * FROM v0 , v0 WHERE v1 = v1 AND v1 = 1;
```

Listing 4: Case Study 4: Bug exists for 1 day

```
01 |CREATE TABLE v0 (v1 DOUBLE CHECK((v1 IN (NULL))),
02 |            v2 UNIQUE AS(v1>v1)) ;
03 |INSERT INTO v0
04 |    VALUES (10) ON CONFLICT DO NOTHING ;
05 |SELECT 10.100000, 10 FROM v0
06 |    CROSS JOIN v0 USING (v1) ;
```

Listing 5: Case Study 4: Bug exists for 1 hour

**Table 6: P-values of SQUIRREL v.s. other fuzzers.** P-value less than 0.05 (shown in green) means the result is statistically significant.

| v.s. Fuzzer | DBMS | Coverage | Syntax | Semantics | Crash | Bug |
|---|---|---|---|---|---|---|
| AFL | SQLite | 0.00609 | 0.00609 | 0.00609 | 0.00609 | 0.00198 |
| | PostgreSQL | 0.00609 | 0.000167 | 0.00225 | - | - |
| | MySQL | 0.00596 | 0.00609 | 0.00609 | - | - |
| SQLsmith | SQLite | 0.00545 | 0.00609 | 0.00609 | 0.00374 | 0.00198 |
| | PostgreSQL | 0.989 | 0.999 | 0.000166 | - | - |
| Angora | SQLite | 0.00609 | 0.00609 | 0.00609 | 0.00374 | 0.00198 |
| GRIMOIRE | SQLite | 0.00583 | 0.00374 | 0.00374 | 0.00374 | 0.00198 |
| QSYM | SQLite | 0.00405 | 0.00405 | 0.00405 | 0.00377 | 0.00198 |
| | PostgreSQL | 0.00583 | 0.000166 | 0.0181 | - | - |
| !semantic | SQLite | 0.00609 | 0.0183 | 0.00609 | 0.00557 | 0.00198 |
| | PostgreSQL | 0.00609 | 0.198 | 0.0181 | - | - |
| | MySQL | 0.00596 | 0.996 | 0.00609 | - | - |
| !feedback | SQLite | 0.00609 | 0.00609 | 0.00609 | 0.00485 | 0.00198 |
| | PostgreSQL | 0.00609 | 0.000167 | 0.000421 | - | - |
| | MySQL | 0.00596 | 0.0718 | 0.338 | - | - |

**Table 7: The absolute number of generated test cases for the evaluated fuzzers in 24 hours.** We categorize them into three groups: one with syntax error, one with syntax correctness and semantic error, and one with syntax correctness and semantic correctness.

| Fuzzer | DBMS | Syntax-error | Semantics-error | Correct | Total |
|---|---|---|---|---|---|
| SQUIRREL | SQLite | 1,627,034 | 10,561,457 | 5,696,308 | 17,884,799 |
| | PostgreSQL | 7,287 | 188,055 | 25,762 | 221,104 |
| | MySQL | 50,728 | 77,750 | 28,314 | 156,792 |
| AFL | SQLite | 29,731,018 | 12,576,971 | 1,496,807 | 43,804,796 |
| | PostgreSQL | 3,604,530 | 245,226 | 57,166 | 3,906,922 |
| | MySQL | 185,102 | 42,332 | 1,478 | 228,912 |
| SQLsmith | SQLite | 20,205,598 | 2,891,250 | 58,351 | 23,155,199 |
| | PostgreSQL | 3,752 | 821,485 | 394 | 825,631 |
| Angora | SQLite | 14,540,639 | 3,866,102 | 654,526 | 19,061,267 |
| GRIMOIRE | SQLite | 4,799,412 | 886,471 | 581,500 | 6,267,383 |
| QSYM | SQLite | 8,652,045 | 2,805,585 | 385,028 | 11,842,658 |
| | PostgreSQL | 78,161 | 1,350 | 1,832 | 81,343 |
| !semantic | SQLite | 24,443,355 | 15,754,468 | 809,209 | 19,008,012 |
| | PostgreSQL | 9,591 | 184,441 | 3,174 | 197,206 |
| | MySQL | 17,140 | 165,175 | 15,370 | 197,685 |
| !feedback | SQLite | 19,415,174 | 8,146,566 | 27,368 | 27,589,108 |
| | PostgreSQL | 29,283 | 26,678 | 174 | 56,135 |
| | MySQL | 66,695 | 26,746 | 19,974 | 113,415 |